\documentclass[aps,prl,twocolumn,showpacs]{revtex4}
\usepackage{amsmath}
\usepackage{graphicx}
\usepackage{units}  
\usepackage{xspace}
\usepackage{subfigure}
\usepackage{hyperref}
\newcommand{\beq}{\begin{equation}}
\newcommand{\eeq}{\end{equation}}
\newcommand{\bea}{\begin{eqnarray}}
\newcommand{\eea}{\end{eqnarray}}

\newcommand{\sx}{\sigma_{ x}}
\newcommand{\sy}{\sigma_{ y}}
\newcommand{\sz}{\sigma_{ z}}

\newcommand{\moire}{Moir\'{e}\xspace}
\newcommand{\Moire}{Moir\'{e}\xspace}
\newcommand{\vf}{v}
 \newcommand{\m}{mv^2}
 
\usepackage[usenames]{color}

\begin{document}
\bibliographystyle{apsrev}

 \title{Zero Energy Modes and  Gate-Tunable  Gap in Graphene on  hexagonal Boron Nitride}

\author{M. Kindermann$^{1}$, Bruno Uchoa$^{2}$ and D. L. Miller$^{3}$}

\affiliation{$^{1}$ School of Physics, Georgia Institute of Technology, Atlanta, Georgia 30332, USA}

\affiliation{$^{2}$ Department of Physics and Astronomy, University of Oklahoma,
Norman, OK 73069, USA}

\affiliation{$^{3}$ National Institute of Standards and Technology, Boulder, CO 80305 USA}

\date{\today}
\begin{abstract}
In this Letter, we derive an effective theory of graphene on a hexagonal Boron Nitride (h-BN)  substrate.  
We show that the h-BN substrate  generically opens a spectral gap in graphene despite the lattice mismatch. The origin of that gap is particularly intuitive in the  regime of strong coupling between graphene and its substrate, when  the low-energy physics  is determined by the topology of a network of   zero energy modes.  For twisted graphene bilayers, where inversion symmetry is present, this network percolates  through the system and the spectrum is gapless.  The breaking of that symmetry by h-BN   causes the  zero energy modes to close into rings. The eigenstates of these rings hybridize into flat bands with  gaps in between. The size of this band gap   can be tuned by a gate voltage and   it can reach the order of magnitude needed to confine electrons at room temperature. 
 \end{abstract}


\maketitle
Graphene is a two-dimensional semimetal with low-energy
excitations that obey the massless Dirac equation \cite{novoselov:nat05,zhang:nat05,neto:rmp09}.  As most applications in electronics require a bandgap, much effort has  been exerted to find ways of inducing a gap in the electronic spectrum of graphene.  One possible route is to use hexagonal boron
nitride (h-BN) substrates \cite{giovannetti:prb07,slawinska:prb10}, which lack sublattice inversion
symmetry.  If inherited by the graphene layer, this broken symmetry  leads to the opening of a gap in the spectrum, which is described by a mass   in the Dirac model.  
First principles calculations for graphene supported by a perfectly lattice matched   h-BN  substrate predicted a bandgap on the order $50 \,{\rm meV}$  \cite{giovannetti:prb07}. Experiments, however, have not observed any clear indication of a gap \cite{xue:nmat11,decker:nal11}.  

\begin{figure}[b] 
\vspace{-0.2cm}
\begin{centering} 
\includegraphics[width=0.65\linewidth]{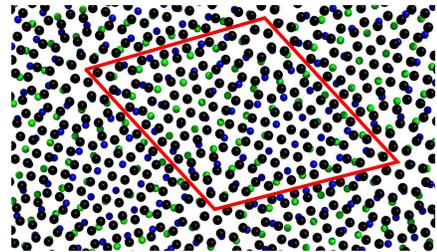}
\par\end{centering}

\caption{{\small Two layer system made of graphene (top layer) and
h-BN (bottom layer) with a lattice mismatch (exaggerated in the figure). Red line: \Moire unit cell of the system.}}
\label{fig1}
\end{figure}

Subsequent theory  \cite{sachs:prb11,ortix:11} identified the reason for this discrepancy: the lattice constants of h-BN and graphene differ by about $1.8\%$ and for  multi-layered h-BN substrates it is energetically unfavorable for graphene and its substrate to conform their lattice constants \cite{sachs:prb11}. The result is \Moire patterns of varying local lattice alignment, as illustrated in Fig.\ \ref{fig1}, which have been observed in scanning tunneling microscopy   \cite{yankowitz:12}. Evidently,   the h-BN substrate breaks sublattice symmetry differently in different regions of the \Moire pattern. Density functional theory  (DFT) calculations have shown a tendency for the sublattice symmetry breaking to be    compensated between different regions of the \Moire pattern such that the   symmetry is  almost restored after spatial average  \cite{sachs:prb11}. This has motivated the proposal of effective Dirac models of graphene on h-BN with a mass term that has vanishing integral, such that sublattice inversion symmetry is restored on (spatial)   average.  On the basis of those models, it was argued that a band gap in graphene on h-BN is absent \cite{sachs:prb11,ortix:11}, consistently with the existing experimental data  \cite{xue:nmat11,decker:nal11}.

In this Letter, we derive an effective theory for graphene on h-BN based on a bilayer model that has been successfully applied to twisted graphene bilayers \cite{lopes:prl07,kindermann:prb11,mele:prb11}.   Our theory is formulated for a single-layer of graphene and it accounts for the coupling to the substrate  by a  mass term and effective potentials that oscillate with the period of the \moire pattern.  We find that graphene supported by a hexagonal substrate  generically develops a gap in the spectrum.  In particular, the emergence of a spectral gap  is \emph{not} precluded by an average sublattice symmetry. A gap is avoided only by additional symmetries, such as in  twisted graphene bilayers, where the Dirac points are topologically protected by a combination of space inversion and time reversal symmetry  \cite{fu:prb07}.  In the regime of strong coupling between graphene and its substrate a particularly intuitive picture of that gap formation emerges:  the oscillatory mass in our effective theory then defines one-dimensional modes  that are topologically protected  and gapless for a large Moir\'e period. In the presence of space inversion symmetry, such as in twisted graphene bilayers, these modes form a network that percolates through the system, corresponding to a metallic state.  When space inversion symmetry is broken, for instance by h-BN, this network breaks up into  isolated rings  of 1D-modes. The states in these rings hybridize exponentially weakly and form narrow bands with large gaps set by the level spacing in the rings.  

Finally, we show that the spectral gap of graphene on h-BN can be controlled by the application of a perpendicular electric field. Using parameters fitted to experiments and DFT calculations, we find that the gap can be tuned to reach the order of magnitude needed to confine charge carriers at room temperature, a key requirement for electronics applications.

\emph{Model:} We base our analysis of graphene with an h-BN substrate on a tight-binding model of two coupled honeycomb lattices with parameters fitted to experiment  \cite{dresselhaus:aip02}.  The two layers have a lattice mismatch of $\delta \approx 1.8\%$ (cf.\ Fig.\ \ref{fig1}) and we allow in addition a rotational misalignment by angle $\theta$.  In the two-layer basis of  graphene and  h-BN, the electronic Hamiltonian  is   
\beq
H= \left(\begin{array}{cc} H_{\rm g} & H_{\rm int} \\ H_{\rm int}^\dagger & H_{\rm BN} \end{array}  \right).
\eeq 
  In the limit $\theta\ll 1$ and $\delta \ll 1$, where a long-wavelength description is appropriate, the isolated graphene layer can be described by a Dirac model Hamiltonian $H_{\rm g}=\vf  {\bf p}\cdot \boldsymbol{\Sigma}$, and  the h-BN layer is similarly described by $H_{\rm BN}=v_{\rm BN}  {\bf p}\cdot \boldsymbol{\Sigma}+\m \sigma_z+V$, where $\vf$ and $v_{\rm BN}$ are the carrier velocities in graphene and h-BN, respectively and  $\mathbf{p}$ is the momentum relative to the Dirac points.  $H_{\rm g}$ and $H_{\rm BN}$ each act on 4 component spinors of the form $(\psi_{{\rm A},+},\psi_{{\rm B},+},\psi_{{\rm A},-},\psi_{{\rm B},-})$, with ${\rm A},\,{\rm B}$  labeling the two different sublattices of the honeycomb lattice and $\pm$ denoting  the two different band structure valleys. $\boldsymbol{\Sigma}=(\tau_z \sx ,\sy)$ is a vector of Pauli matrices acting on the A/B sublattice basis (through the pseudospin $\boldsymbol{\sigma}$) and the valley spin  ($\boldsymbol{\tau}$). The mass  $m$ and the interlayer bias $V$ account for the different on-site potentials of the Boron (the A sublattice of h-BN) and the Nitrogen (B-)atoms. DFT calculations indicate  $\m\approx 2.3 \,{\rm eV}$ and $V\approx 0.8 \,{\rm eV}$ \cite{slawinska:prb10}. The bias $V$ can be tuned by a perpendicular electric field.  

Following Ref.\ \cite{mele:prb11}, we take the interlayer coupling of Eq. (1) at long-wavelengths to be of the form    \cite{lopes:prl07,mele:prb11}
\beq \label{t}
H_{\rm int}=\!\frac{\gamma}{3}  \sum_{n}\!e^{i \tau_z \delta\!\boldsymbol{K}_{n}\cdot\boldsymbol{r}} \left(\begin{array}{cc} 1 & \zeta \, e^{i\tau_z\phi_n} \\ \zeta \,  e^{-i\tau_z\phi_n}& 1 \end{array}  \right).
\eeq
Here, $H_{\rm int}$ is written explicitly in the ${\rm A}$, ${\rm B}$ sublattice basis, $\gamma\approx 0.3$ eV is the hopping energy to the substrate,  $\zeta$ parametrizes a sublattice asymmetry due to structural differences between different regions of the \Moire pattern   \cite{mele:prb11}, and $\phi_n=2\pi n /3$ is a phase that depends on the index $n=0,1,2$, which labels the three  corners of the graphene Brillouin zone corresponding to a given valley. Those points have  wavevectors  $ \boldsymbol{K}_{n} = R(\phi_n+\pi/6) \hat{a} \,4\pi/3\sqrt{3a}$, where  $\hat{a}$ is a unit vector along an A-B bond,   $a$ the bond length, and $R(\varphi)$ is a rotation by angle $\varphi$. The wavevectors $\delta\!\boldsymbol{K}_{n}$ are the differences of  $ \boldsymbol{K}_{n} $ and their counterparts  in the closest valley of the h-BN layer. They are shorter than $ \boldsymbol{K}_{n} $ by a  factor $ \delta \!  {K}_{n} / {K}_{n} = \sqrt{\delta^2-2(1+\delta)(\cos\theta-1)}/(1+\delta)$ and rotated  with respect to   $ \boldsymbol{K}_{n} $ by  angle $\Phi = \arctan \left[ \sin\theta/( 1+\delta-\cos\theta)\right]$. 
We neglect commensuration effects, which are  small for   $\delta, \theta \ll 1$ \cite{mele:prb10}.

{\em Effective single-layer theory:}
Integrating out the electrons in the h-BN layer, we arrive at an effective Hamiltonian $H_{\rm g}^{\rm eff}(\omega)=H_{\rm g}+\delta\!H_{\rm g}^{\rm eff}(\omega)$ for the graphene layer with  (we set $\hbar =1$) 
\beq \label{Heff}
\delta\!H_{\rm g}^{\rm eff}(\omega)= H_{\rm int}(\omega-H_{\rm BN})^{-1}H_{\rm int}.
\eeq
The mass term $m$ dominates the Hamiltonian $H_{\rm BN}$ for all wavevectors where the employed Dirac model holds. At those momenta, $p \ll K_n$,   we may set $v_{\rm BN}=0$ to a good approximation, resulting in an effective Hamiltonian which is local in space \footnote{In contrast to twisted graphene bilayers, where that statement does not generally hold.},
\beq \label{local}
\delta\!H_{\rm g}^{\rm eff}=  \frac{1}{\omega-V+\m}\, H_{\rm int}\left(\begin{array}{cc}  \eta & 0 \\ 0 & 1 \end{array} \right) H_{\rm int}, 
\eeq
where  $\eta$ parametrizes the inversion symmetry breaking through the h-BN substrate, 
 \beq
 \eta=(\omega-V+\m)/(\omega-V-\m).
 \eeq 
The effective Hamiltonian (\ref{local}) can be parametrized in terms  of effective  potentials that oscillate in space with the periodicity of the bilayer \Moire pattern, 
\beq \label{para}
\delta\!H_{\rm g}^{\rm eff}= V^{\rm eff}({\bf r})+\nu \vf e\boldsymbol{\Sigma} \cdot \boldsymbol{A}^{\rm eff}({\bf r})+m^{\rm eff}({\bf r}) \vf^2 \sz .
\eeq
For perfect rotational alignment,  $\theta=\Phi=0$, the effective vector potential $\boldsymbol{A}^{\rm eff}$ may be gauged away.  In the more general case $\Phi\neq0$  the vector potential generates a pseudo-magnetic field and it satisfies the Coulomb gauge condition $\boldsymbol{\nabla}\cdot  \boldsymbol{A}^{\rm eff}=0$ at $\Phi =\pi/2$. The  mass term $m^{\rm eff}(\mathbf{r})$  breaks the sublattice exchange symmetry locally and it opens a local gap in the spectrum wherever it exceeds $1/vL$, such that the wavefunctions are localized on the length scale $L$ of the \Moire pattern. 

A global gap in the spectrum is nevertheless precluded when  the effective theory Eq.\ (\ref{local}) is   invariant  under $P=\sigma_x\tau_x R(\pi)$ \footnote{Provided the effective potentials are not too large [22]},  that is  sublattice exchange $\sx$ coupled with point reflection $R(\pi)$ and valley exchange $\tau_x$, such that the underlying lattice model has inversion and time reversal symmetry,  as for twisted graphene bilayers. In the following, we analyze the spectral gap of graphene on h-BN  in the absence of that symmetry, which is explicitly broken by the inequivalence of the B and N sites. 

\begin{figure}[t] 
\begin{centering}
\includegraphics[scale=0.18]{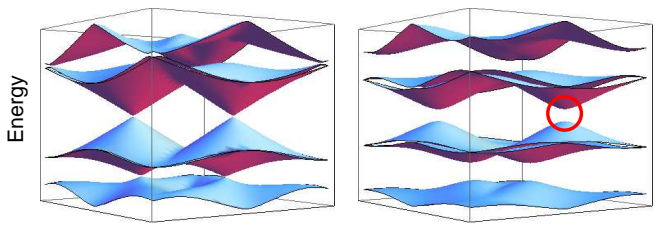}

\par\end{centering}
\caption{{\small  Energy spectra in the \Moire Brillouin zone for strong coupling   $\varepsilon = L\gamma^2/[(V-mv^2)v]\gg1$.  (Left)   $\eta=1$, as in twisted graphene bilayers (corresponding to the triangular network of zero energy modes shown in Fig. \ref{fig3} a), with  a metallic spectrum. (Right)  $\zeta=1$ and $\eta = -0.5$    (corresponding to   isolated rings of low-energy modes, as shown in Fig. \ref{fig3} b-d), with  narrow bands separated by large gaps (red circle). }}
\label{fig2}
\end{figure}

{\em Perturbation theory:} We start our analysis of the effective theory Eq.\ (\ref{local}) by a perturbative calculation, valid for  weak coupling  $\gamma^2/|\omega-V\pm \m| \ll  \vf \delta K$. For a lattice mismatch of $\delta\sim 1.8\%$ one has $\vf \delta K \gtrsim 0.22 \, {\rm eV}$, where the lower bound corresponds to $\theta=0$. Since  $\gamma \approx 0.3$ eV and in the absence of an external bias $|V\pm m|\gtrsim 1.5$ eV,   the system is well in this perturbative regime at low energies ($|\omega|\ll |V\pm m|$ eV).   

When $\eta=1$ the model has inversion symmetry $P$ in addition to time reversal invariance and topological arguments  \cite{fu:prb07} require the presence of at least two Dirac points (or arcs)    \footnote{The argument requires that the product of the parity eigenvalues at the three M-points and the  $\Gamma$-point  of the \moire Brillouin zone   is $-1$. This condition is satisfied throughout the perturbative regime, where the effective potentials are not able to move states at $M$ or $\Gamma$ through the Fermi level \cite{fu:prb07}.\label{foot}}  in the \Moire Brillouin zone. This is the situation in twisted graphene bilayers.

When the  sublattice symmetry of the substrate is broken  ($\eta \neq  1$) such as through h-BN,   a gap is not precluded by symmetry anymore.  In the case  $\zeta=1$ it turns out that the spatial average $\langle m^{\rm eff}\rangle =\int{\rm d}\mathbf{r}\, m^{\rm eff}(\mathbf{r})$ vanishes and correspondingly no gap is found to leading order   in the effective potentials.  Unlike previously assumed \cite{sachs:prb11,ortix:11}, this restoration of a  symmetry on (spatial) average, however, does not suppress the gap in the spectrum entirely. A gap does appear at third order in $\delta\!H_{\rm g}^{\rm eff}$  \footnote{In  case  the two layers are commensurate there will be a contribution to the gap that is quadratic in $\gamma$ and that is induced by a momentum-conserving term in the effective interlayer Hamiltonian of the form first pointed out by Mele \cite{mele:prb10} and neglected in our above model.}:
\beq \label{cubic}
\Delta=\left| \eta(1-\eta)(2 \cos 2\Phi-1) \frac{\gamma^6}{81v^2 \delta K^2 (V-\m)^3}\right|.
\eeq

In the more general situation, when $\zeta\neq 1$,  the spatial average  of the mass term $\langle m^{\rm eff}\rangle$ is non-zero, and the band gap appears already at leading order in perturbation,
\beq \label{firstorder}
\Delta= \left|(1-\eta)(1-\zeta^2)\frac{\gamma^2}{3(V-\m)}  \right|.
\eeq
A recent  DFT calculation  \cite{sachs:prb11} predicted $\Delta \approx 4\, {\rm meV}$ in the absence of an external perpendicular field, when $\eta \approx -0.5$. Assuming  $\gamma=0.3 \, {\rm eV}$, we estimate  $|1-\zeta^2| \approx 0.14$ from Eq.\ (\ref{firstorder}). The relative magnitude of the local gaps in various \Moire regions found in the DFT calculation \cite{sachs:prb11} indicate   $\zeta>1$, so we conclude that $\zeta \approx 1.07$. A direct fit to the local gaps of Ref.\  \cite{sachs:prb11} in the AA-, AB-, and BA-stacked regions, respectively, yields $\zeta=1.19$, $\eta=-0.72$, and $\gamma=0.25 \, {\rm eV}$. One possible reason for the discrepancy is higher Fourier harmonics of $\delta H^{\rm eff}$ that we neglect.  

 Eq.\ (\ref{firstorder}) predicts that the spectral gap of graphene on h-BN can be substantially increased by application of a perpendicular electric field that decreases the B/N on-site energies $V\pm \m$.  As the gap increases, perturbation theory eventually breaks down, and the gap has a crossover to  a nonperturbative regime.  

\begin{figure} 
\begin{centering}
\includegraphics[scale=0.33]{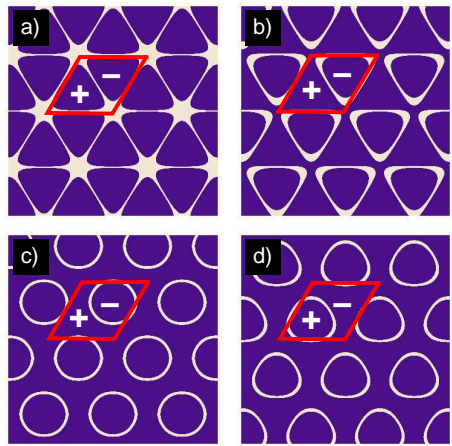}
\par\end{centering}

\caption{{\small Spatial dependence of the mass term resulting from the local lattice misalignment in a bilayer as shown in Fig.\ \ref{fig1}. The red line indicates the \Moire  unit cell, and the $\pm$ signs specify the sign of the mass. a) $\eta=1$: metallic state, with percolating zeros of $m^{\rm eff}$; b) $\eta=0.9$; c) $\eta=0.5$;
d) $\eta=-0.5$: the network of zeros breaks up into  rings. In the strong coupling regime, those rings contain fully localized states.}}
\label{fig3}
\end{figure}

{\em Nonperturbative regime:} In the nonperturbative limit $\gamma^2/|\omega-V\pm \m| \gg  \vf \delta K$
the spectrum is gapped locally in regions where $m^{\rm eff}(\mathbf{r})\neq 0$. The low-energy physics of the system is then dominated by  one-dimensional modes along the zeros of $m^{\rm eff}$ \cite{martin:prl08,semenoff:prl08}.  In the absence of intervalley scattering and in the limit of a large \Moire size $L\to\infty$ these modes are guaranteed to be gapless  by topological arguments \cite{volovik:03}: the lines with $m^{\rm eff}=0$
  separate  regions with effective
masses of opposite signs, as indicated by the {}``+'' and {}``-''
signs   in Fig. \ref{fig3}. 
The massive, single-valley Dirac   Hamiltonian   $H=\mathbf{g}(\mathbf{k})\cdot\boldsymbol{\sigma}$,
where $\mathbf{g}=v(k_{x},k_{y},m^{\rm eff}v)$ and $\boldsymbol{\sigma}=(\sigma_{x},\sigma_{y},\sigma_{z})$, has a topological
charge  $N_{3}= \int\!\mbox{d}k_{x}\mbox{d}k_{y}\,\hat{\mathbf{g}}\!\cdot\!\left(\partial_{k_{x}}\hat{\mathbf{g}}\times\partial_{k_{y}}\hat{\mathbf{g}}\right)/4\pi=m^{\rm eff}/(2|m^{\rm eff}|)$ associated with it (here, $\mathbf{\hat{g}}=\mathbf{g}/|\mathbf{g}|$). The difference of the charges $N_{3}^{+}$ and $N_{3}^{-}$ to both  sides of a line $m^{\rm eff}=0$ enforces $|N_{3}^{+}-N_{3}^{-}|=1$ zero modes per valley \cite{volovik:03}. 
For  rotational alignment $\Phi=0$, when ${\bf A}^{\rm eff}$ is a pure gauge and for strong screening of the scalar potential $V^{\rm eff}$,
 the qualitative low-energy physics thus  is  determined entirely by the topology of the zeros of  $m^{\rm eff}$.  

When $\eta=1$, such as in  twisted graphene bilayers,  the zeros of  $m^{\rm eff}(\mathbf{r})$ form a triangular network that percolates through the entire system, as seen in Fig.\ \ref{fig3}a. The expected pairs of  zero modes  along these lines provide an intuitive explanation for the metallic state required by the topological arguments quoted earlier \cite{fu:prb07}.
  
For graphene on h-BN with $\eta\neq1$ on the other hand,  this triangular network of   zero energy modes   breaks up into {\it isolated} rings (cf.\ Fig.\  \ref{fig3} b-d). The  low-energy modes confined to the rings of $m^{\rm eff}(\mathbf{r})=0$ form discrete states with an energy separation set by the circumference $C\simeq L$ of those rings. The states hosted by neighboring rings have an overlap that is exponentially small in $L \gamma^2/(V-\m) \vf $ and form correspondingly narrow (valley) pairs of bands with gaps $\Delta \approx 2\pi \vf/C$ in between. Evidently, in this strong coupling regime, a gap of that order of magnitude will appear regardless of whether $\langle m^{\rm eff}\rangle$ vanishes or not. This gives a physical reason why an average sublattice symmetry cannot prevent the opening of a gap in the spectrum.

  \begin{figure} 
\begin{centering}
\includegraphics[scale=0.35]{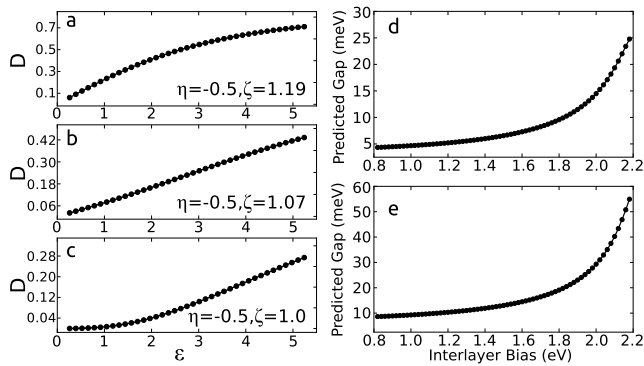}
\par\end{centering}

\caption{{\small Left panels: numerical scaling of the gap $\Delta \propto D(\varepsilon,\eta,\zeta)$ as a function of $\varepsilon = L \gamma^2/[(V-mv^2)v]$ for realistic parameter values at $\theta=0$. In weak coupling ($\varepsilon\ll 1$), the predicted crossover from linear scaling ($D \propto \varepsilon$) at $\zeta\neq 1$ (a,b),  [Eq.\ (\ref{firstorder})],  to cubic  ($D \propto \varepsilon^3$) at $\zeta=1$ (c) [Eq.\ (\ref{cubic})], is confirmed.  Right: Predicted gap as a function of the interlayer bias $V$ (tunable by a perpendicular electric field) for parameter values from two  fits to DFT: $\gamma=0.3 \, {\rm eV}$, $\eta = -0.5$, $\zeta = 1.07$ (d) and $\gamma=0.25 \, {\rm eV}$,  $\eta=-0.72$, $\zeta = 1.19$ (e) (see text).}}
\label{fig4}
\end{figure}

 For graphene on h-BN  the above nonperturbative considerations do not strictly apply: there is no $V$ such that the limit    $\gamma^2/|\omega-V\pm\m| \gg  \vf \delta K$ is reached for all energies $\omega$ inside  the predicted gap of order $2\pi \vf/C  \approx  \vf \delta K \geq220\, {\rm meV}$. We thus next perform numerical calculations that use a tight-binding model. Computations for the true size of the unit cell  at $\theta=0$  are challenging. We therefore exploit the scale invariance of our theory, expressing
 \beq
 \Delta= D( \gamma^2/(V-\m)\vf\delta K,\eta,\zeta)\times  {\vf} \delta K
 \eeq
  in terms of a  function $D(\varepsilon,\eta,\zeta)$ that may be evaluated for smaller unit cells    \footnote{Provided that the rescaled potentials $\delta H^{\rm eff} L/l$ remain smaller than the band cut-off, where the Dirac model ceases to apply.}.
 The scaling parameter $\varepsilon= \gamma^2/[(V-\m)\vf\delta K]$ separates the weak ($\varepsilon\ll1$) from the strong ($\varepsilon\gg1$) coupling regime.   

In Fig. \ref{fig3}a-c, we plot the scaling function $D$  found from tight-binding calculations on a unit cell containing 512 atoms for typical parameters. When $\zeta=1$, the scaling of the gap is cubic in $\varepsilon$ for  weak coupling, crossing over to linear ($\Delta\propto \varepsilon$) behavior for $\zeta\neq 1$, in agreement with the perturbative analysis of  Eqs.\ (\ref{cubic}) and (\ref{firstorder}).    Fig.\ \ref{fig3}d-e shows the gap as a function of  $V$ for  parameters taken from the two above fits to DFT data   \footnote{The plot neglects the $\omega$-dependence of the effective potentials, which is a good approximation in the range of $V$ shown in Fig.\ \ref{fig3}.}.  Despite the uncertainty of the parameters entering our model these calculations clearly suggest that it is possible to induce gaps in graphene on h-BN on the order of room temperature.

 {\em Conclusions:} We have derived a low-energy theory for graphene on hexagonal substrates. Our theory demonstrates that  a h-BN  substrate opens a gap in the spectrum of graphene through a breaking of  inversion symmetry even when the sublattice symmetry of graphene is restored on (spatial) average. We moreover have shown that perpendicular electric fields may be used to enhance the predicted gaps up to the  scale of room temperature.
 
 We thank A.\ H.\ Castro Neto, P.\ M.\ Goldbart, P. Jarillo-Herrero and E.\ J.\ Mele for discussions and gratefully acknowledge  financial support from the NSF  (DMR-1055799 and DMR-0820382). 
Contribution of an agency of the U.S. government, not subject to copyright.

 \end{document}